\documentclass{article}
\usepackage{amsfonts}
\usepackage{amsmath}

\newtheorem{theorem}{Theorem}

\newtheorem{definition}[theorem]{Definition}

\newtheorem{proposition}[theorem]{Proposition}
\newtheorem{remark}[theorem]{Remark}

\input{tcilatex}
\begin{document}

\title{A new approach to electromagnetism in anisotropic spaces}
\author{Nicoleta VOICU \\
%EndAName
Transilvania University, Brasov, Romania \and Sergey SIPAROV \\
%EndAName
State University of Civil Aviation, St. Petersburg, Russia}
\maketitle

\begin{abstract}
Anisotropy of a space naturally leads to direction dependent electromagnetic
tensors and electromagnetic potentials. Starting from this idea and using
variational approaches and exterior derivative formalism, we extend some of
the classical equations of electromagnetism to anisotropic (Finslerian)
spaces. The results differ from the ones obtained by means of the known
approach in \ \cite{Lagrange}, \cite{Miron-Rad}.
\end{abstract}



\section{Introduction}

\qquad In anisotropic spaces, where the fundamental metric tensor depends on
the directional variables, the electromagnetic-type tensor $F$ and
accordingly, the electromagnetic potential $A,$ may also depend on these.

Starting from this idea, we propose a generalization of the electromagnetic
tensor, of the notion of current and of the corresponding Maxwell equations
- based on variational methods and exterior derivative formalism. We chose
those anisotropic spaces which provide the simplest equations, namely,
Finslerian ones. A similar approach for a particular class of Finsler spaces
was already considered by the authors in \cite{Cairo08}.

When dealing with the equations of electromagnetism, one can either: 1)
consider as a fundamental object the electromagnetic tensor $F$ satisfying
the homogeneous Maxwell equations (written in a condensed manner as $dF=0$)
and deduce by Poincar\'{e} lemma the existence of a potential 1-form $A$
such that $F=dA,$ or, conversely: 2) consider the potential 1-form $A$ as a
fundamental object, and define the electromagnetic tensor as its exterior
derivative - thus getting the homogeneous Maxwell equations as identities.

In order to realize how the electromagnetic tensor and Lorentz force might
look like in anisotropic spaces, it appeared as more convenient to use for
the beginning the second approach, and then point out (Theorem \ref{th1})
that using the first one, we are led to similar results.

The theory we are going to develop stems from considering\emph{\ }a \textit{%
direction dependent potential} 4-covector field $A=A(x,y),$ (where, $%
x=(x^{i})$ are the space-time coordinates and $y=(y^{i})$, the directional
ones) as arising from a Lagrangian. Namely, it appears as reasonable to
consider the following Lagrangian $\mathcal{L}$, providing the Lorentz force
in Finsler spaces: 
\begin{equation*}
\mathcal{L}=\dfrac{1}{2}g_{ij}(x,y)y^{i}y^{j}+\dfrac{q}{c}L_{1}(x,y),~y^{i}=%
\dot{x}^{i},
\end{equation*}%
where $L_{1}$ is a 1-homogeneous function in $y,$ $(L_{1}(x,\lambda
y)=\lambda L_{1}(x,y),\lambda \in \mathbb{R})$ and $g$ is the Finslerian
metric tensor. Then, the Liouville (canonical) 1-form 
\begin{equation*}
\theta =\dfrac{\partial \mathcal{L}}{\partial y^{i}}dx^{i}
\end{equation*}%
and the Poincar\'{e} 2-form $\omega =d\theta $ (where $d$ denotes exterior
derivative) attached to $\mathcal{L}$ carry information on both the metric
of the space and on electromagnetic properties. The potential 1-form $A$ can
be defined as $A=\theta -y_{i}dx^{i},$ which is,%
\begin{equation*}
A=A_{i}(x,y)dx^{i},~\ A_{i}=\dfrac{\partial L_{1}}{\partial y^{i}}.
\end{equation*}%
and the electromagnetic tensor, as $F=\omega -g_{ij}\delta y^{j}\wedge
dx^{i},$ which is nothing but the exterior derivative of $A:$ 
\begin{equation*}
F=dA=\dfrac{1}{2}(A_{j|i}-A_{i|j})dx^{i}\wedge dx^{j}-\dfrac{\partial A_{i}}{%
\partial y^{a}}dx^{i}\wedge \delta y^{a},
\end{equation*}%
(where bars denote Chern type covariant derivatives)

The equations of motion of charged particles are then 
\begin{equation*}
\dfrac{\delta y^{i}}{dt}=\dfrac{q}{c}F_{~j}^{i}y^{j}+\dfrac{q}{c}\tilde{F}%
_{~a}^{i}\dfrac{\delta y^{a}}{dt},~\ y^{i}=\dot{x}^{i}
\end{equation*}%
where $\dfrac{\delta y^{i}}{dt}=\dfrac{dy^{i}}{dt}+\Gamma
_{~jk}^{i}(x,y)y^{j}y^{k},$ and $F_{ih}=A_{j|i}-A_{i|j},~\ \tilde{F}_{ia}=-%
\dfrac{\partial A_{i}}{\partial y^{a}}$ are the components of the
electromagnetic tensor above defined.

Maxwell equations in Finsler spaces are obtained as 
\begin{eqnarray}
dF &=&0  \label{homog_Max} \\
(d\ast F) &=&\dfrac{\beta }{4\alpha }(\ast \mathcal{J)}.  \label{current}
\end{eqnarray}%
where $\ast $ denotes Hodge star operator, and the \textit{current} $%
\mathcal{J}$ is a vector field on $TM,$ and $\alpha ,\beta $ are constants.
In order to obtain the expression for currents, we also perform a
variational approach (with an integrand defined on a domain in $TM$).

From a physical point of view, we notice the appearance of an additional
term (reminding inertial forces) in the expression of Lorentz force (\ref%
{Lorentz1}), as well as the appearance of a correction to the usual
expression of currents, (\ref{new-current}).

The generalized current $\mathcal{J}=J^{i}\delta _{i}+\tilde{J}^{a}\dot{%
\partial}_{a}$ obeys the continuity equation $div\mathcal{J}=0.$ The
horizontal component $J^{i}$ is equal to the regular current plus a
correction due to anisotropy, while the vertical one $\tilde{J}^{a}$ plays
the role of compensating quantity so as to have the continuity equation
satisfied.

\section{Pseudo-Finsler spaces}

Let $M$ be a 4-dimensional differentiable manifold of class $\mathcal{C}%
^{\infty },$ thought of as spacetime manifold, $(TM,\pi ,M)$ its tangent
bundle and $(x^{i},y^{i})_{i=\overline{1,4}}$ the coordinates in a local
chart on $TM.$ By "smooth" we shall always mean $\mathcal{C}^{\infty }$%
-differentiable. Also, we denote partial derivation with respect to $x^{i}$
by \thinspace $_{,i}$ and partial derivation with respect to $y^{i},$ by a
dot: $_{\cdot i}.$

A \textit{pseudo-Finslerian function} on $M,$ is a function $\mathcal{F}%
:TM\rightarrow \mathbb{R}$ with the properties, \cite{Szilasi}:

\begin{enumerate}
\item $\mathcal{F=F}(x,y)$ is smooth for $y\not=0;$

\item $\mathcal{F}$ is positive homogeneous of degree 1, i.e., $\mathcal{F}%
(x,\lambda y)=\lambda \mathcal{F}(x,y)$ for all $\lambda >0;$

\item The \textit{pseudo-Finslerian metric tensor}:\textit{\ } 
\begin{equation}
g_{ij}(x,y)=\dfrac{1}{2}\dfrac{\partial ^{2}\mathcal{F}^{2}}{\partial
y^{i}\partial y^{j}}  \label{Finsler_metric}
\end{equation}%
is nondegenerate: $\det (g_{ij}(x,y))\not=0,~\forall x\in M,$ $y\in
T_{x}M\backslash \{0\}.$
\end{enumerate}

Particularly, we shall consider that the metric has signature $(+,-,-,-).$

\bigskip

The equations of geodesics $s\mapsto (x^{i}(s))$ of a Finsler space $(M,%
\mathcal{F})$ are 
\begin{equation*}
\dfrac{dy^{i}}{ds}+2G^{i}(x,y)=0,~\ y^{i}=\dot{x}^{i}.
\end{equation*}

These equations give rise to the \textit{Cartan nonlinear connection }on $TM,
$ of local coefficients%
\begin{equation*}
N_{~j}^{i}=\dfrac{\partial G^{i}}{\partial y^{j}}.
\end{equation*}

Let%
\begin{equation*}
\delta _{i}=\dfrac{\partial }{\partial x^{i}}-N_{~i}^{a}\dfrac{\partial }{%
\partial y^{a}},~\ \ \dot{\partial}_{a}=\dfrac{\partial }{\partial y^{a}}
\end{equation*}%
be the adapted basis corresponding to the Cartan nonlinear connection and 
\begin{equation*}
(dx^{i},~\delta y^{a}=dy^{a}+N_{~i}^{a}dx^{i}),
\end{equation*}%
its dual basis. We will also denote by semicolons adapted derivatives:%
\begin{equation*}
f_{;i}=\delta _{i}f,~\ \ \forall f\in \mathcal{F}(TM).
\end{equation*}

\bigskip

Any vector field $V$ on $TM$ can be written as $V=V^{i}\delta _{i}+\tilde{V}%
^{a}\dot{\partial}_{a};$ the component $hV=V^{i}\delta _{i}$ is a vector
field, called the \textit{horizontal }component of $V,$ while $vV=\tilde{V}%
^{a}\dot{\partial}_{a}$ is its \textit{vertical }component. Similarly, a
1-form $\omega $ on $TM$ can be decomposed as $\omega =\omega _{i}dx^{i}+%
\tilde{\omega}_{a}\delta y^{a},$ with $h\omega =\omega _{i}dx^{i}$ called
the \textit{horizontal} component, and $v\omega =\tilde{\omega}_{a}\delta
y^{a}$ the \textit{vertical }one\textit{.}

In terms of the Cartan nonlinear connection, the divergence of a vector
field $V=V^{i}\delta _{i}+\tilde{V}^{a}\dot{\partial}_{a}\in \mathcal{X}(TM),
$ is obtained, \cite{horiz-Laplacian} (from $d(\ast V^{\flat })=divV\sqrt{G}%
dx^{1}\wedge ...\wedge \delta y^{4}$) as%
\begin{equation*}
divV=\dfrac{1}{\sqrt{G}}\delta _{i}(\tilde{V}^{i}\sqrt{G})-N_{~i\cdot
a}^{a}V^{i}+\dfrac{1}{\sqrt{G}}(\tilde{V}^{a}\sqrt{G})_{\cdot a}.
\end{equation*}%
where $G=\det (G_{\alpha \beta })$ is the determinant of the Sasaki lift of $%
g_{ij}:$ 
\begin{equation}
G_{\alpha \beta }(x,y)=g_{ij}(x,y)dx^{i}\otimes dx^{j}+g_{ij}(x,y)\delta
y^{i}\otimes \delta y^{j}.  \label{Sasaki}
\end{equation}%
\bigskip 

Also, it is convenient to express the electromagnetic tensor in terms of the 
\textit{Chern linear connection} $C\Gamma (N)=(L_{~jk}^{i},0)$ of local
coefficients: 
\begin{equation*}
L_{~jk}^{i}=\dfrac{1}{2}g^{ih}(g_{hj;k}+g_{hk;j}-g_{jk;h}).
\end{equation*}%
We denote by $_{|i}$ and $_{\cdot i}$ the corresponding covariant
derivations 
\begin{equation*}
X_{~|i}^{j}=\delta _{i}X^{j}+L_{~ki}^{j}X^{i},~\ \ X_{~\cdot i}^{j}=\dfrac{%
\partial X^{j}}{\partial y^{i}},
\end{equation*}%
(where $X^{j}$ are local coordinates of a vector field $X$ on $TM$). Then,
we have%
\begin{equation*}
g_{ij|k}=0
\end{equation*}%
(the connection is h-metric). Also, the Chern connection above, there hold
the equalities%
\begin{equation}
y_{i|j}=0.  \label{deflection}
\end{equation}

\bigskip

The \textit{vertical endomorphism} or \textit{almost tangent structure} of $%
TTM,$ \cite{Szilasi}, is the $\mathcal{F}(TM)$-linear function $\mathbf{J}%
:TTM\rightarrow TTM,$ which acts on the elements of the adapted basis as 
\begin{equation*}
\mathbf{J}(\delta _{i})=\dot{\partial}_{i},~\ \mathbf{J}(\dot{\partial}%
_{i})=0,
\end{equation*}%
where $\mathcal{F}(TM)$ denotes the set of smooth real valued functions
defined on $TM.$

For a smooth function $f:TM\rightarrow \mathbb{R},$ the \textit{vertical
differential} $d_{\mathbf{J}}f,$ \cite{Szilasi}, is defined by $d_{\mathbf{J}%
}f=df\circ \mathbf{J};$ in local writing, 
\begin{equation*}
d_{\mathbf{J}}f=\dfrac{\partial f}{\partial y^{j}}dx^{j}.
\end{equation*}

Whenever convenient or necessary to make a clear distinction, we shall
denote by $i,j,k,...$ indices corresponding to horizontal geometrical
objects, and by $a,b,c,...$ indices corresponding to vertical ones.

\section{Direction dependent electromagnetic potential. Electromagnetic
tensor}

In anisotropic spaces and particularly, in Finsler spaces, the components of
an electromagnetic-type tensor $F_{ij},$\ $F_{~j}^{i},$\ $F^{ij}$ and
accordingly, of the electromagnetic potential 1-form\ $A$ basically depend
on the directional variables $y^{i},~\ i=1,...,4.$ 

In order to make sure of this, let us notice the following simple example.
In isotropic (pseudo-Riemannian) spaces with vanishing Ricci tensor, under
some simplifying assumptions, the components of the free electromagnetic
potential 4-vector $A^{i}=A^{i}(x)$ obey Maxwell- de Rham equations, \cite%
{Rham1}: 
\begin{equation*}
g^{ij}(x)\nabla _{i}\nabla _{j}(A^{k})=0,
\end{equation*}%
where $\nabla _{k}=\nabla _{\tfrac{\partial }{\partial x^{k}}}$ denotes
covariant derivative with respect to Levi-Civita connection.

When passing to \textit{anisotropic} spaces with metric $g_{ij}=g_{ij}(x,y),$
the solution of such an equation would generally depend on the directional
variables $y^{i}$ (not to mention that the equation itself could become more
complicated). So, it is meaningful to consider that \textit{the potential
4-vector (and, accordingly, the corresponding 1-form }$A$\textit{) also
depends on the directional variables }$y=(y^{i})$.

Let us define this potential.

In \textit{isotropic} (pseudo-Riemannian) spaces, the Lagrangian providing
Lorentz force is $L(x,y)=\dfrac{1}{2}g_{ij}(x)y^{i}y^{j}+\dfrac{q}{c}%
A_{i}(x)y^{i},$\ $y^{i}=\dot{x}^{i},$ where $q$\ is the electric charge, and 
$A_{i}(x)$ are the covariant components of the 4-vector potential.

For Finsler spaces, let us relax the condition that $L_{1}=A_{i}y^{i}$
should be a linear function of $y$: namely, instead of linearity, we impose
that $L_{1}$ should be just 1-homogeneous in $y,$ which is equivalent to%
\begin{equation*}
\dfrac{\partial L_{1}}{\partial y^{i}}y^{i}=L_{1}.
\end{equation*}

From a physical point of view, this means that we will allow the potential $A%
\emph{\ }$to depend on the directional variable $y$, but not on the
magnitude of $y$. That is, in order to obtain the expression for Lorentz
force in Finsler spaces, we consider the Lagrangian%
\begin{equation}
\mathcal{L}=\dfrac{1}{2}g_{ij}(x,y)y^{i}y^{j}+\dfrac{q}{c}L_{1},
\label{Lorentz_Lagrangian}
\end{equation}%
where $L_{1}=L_{1}(x,y)$ is a scalar function which is 1-homogeneous in the
directional variables.

Let 
\begin{equation*}
\theta =d_{\mathbf{J}}\mathcal{L},
\end{equation*}%
be the \textit{Liouville (canonical) 1-form} attached to $\mathcal{L}.$ In
local coordinates, 
\begin{equation*}
\theta =\dfrac{\partial \mathcal{L}}{\partial y^{i}}dx^{i}=(y_{i}+\dfrac{q}{c%
}\dfrac{\partial L_{1}}{\partial y^{i}})dx^{i}.
\end{equation*}

\begin{definition}
We call \textit{potential 1-form} $A,$ the 1-form given by 
\begin{equation*}
\dfrac qcA=\theta -y_idx^i.
\end{equation*}
\end{definition}

In local writing, 
\begin{equation}
A=A_{i}(x,y)dx^{i},~\ \ \ \ \ A_{i}(x,y)=\dfrac{\partial L_{1}}{\partial
y^{i}}.  \label{def_A}
\end{equation}%
By the 1-homogeneity of $L_{1},$ there holds $L_{1}=A_{j}(x,y)y^{j},$ hence
it makes sense

\begin{definition}
We call Lorentz force Lagrangian in the Finsler space $(M,\mathcal{F})$, the
following function%
\begin{equation}
\mathcal{L}(x,y)=\dfrac{1}{2}g_{ij}(x,y)y^{i}y^{j}+\dfrac{q}{c}%
A_{i}(x,y)y^{i}.  \label{L}
\end{equation}%
The quantities $A_{j}=A_{j}(x,y),$ thus, become the components of a
direction dependent electromagnetic potential. They have the property%
\begin{equation}
A_{i\cdot k}y^{k}=0;~A_{i\cdot k}y^{i}=0.  \label{A}
\end{equation}
\end{definition}

\begin{remark}
In \textit{isotropic} spaces, there exists only one potential 4-vector
providing a given $L_{1}=A_{i}(x)y^{i}$ (which is $A_{i}=\dfrac{\partial
L_{1}}{\partial y^{i}}$). In anisotropic spaces, there exist infinitely many
covector fields $A_{i}=A_{i}(x,y)$ with $A_{i}y^{i}=L_{1}$ for a fixed $%
L_{1}.$ Among them, (\ref{def_A}) is the one which gives the simplest
equations of motion.
\end{remark}

\bigskip

Taking (\ref{deflection}) into account, the exterior derivative of the
1-form $\theta $ yields the following \textit{gravito-electromagnetic\ 2-form%
}: 
\begin{equation}
\omega =d\theta =\dfrac{1}{2}(A_{j|i}-A_{i|j})dx^{i}\wedge
dx^{j}-(g_{ia}+A_{i\cdot a})dx^{i}\wedge \delta y^{a},  \label{omega}
\end{equation}%
which contains information both on the metric structure of the space and on
the electromagnetic field.

\bigskip

\textbf{Particular cases: }

\begin{enumerate}
\item If $A_{i}=A_{i}(x)$ is \textit{isotropic,} then $\tilde{F}_{ia}=0$ and
the 2-form $\omega $ is simply 
\begin{equation*}
\omega =\dfrac{1}{2}(A_{j|i}-A_{i|j})dx^{i}\wedge dx^{j}-g_{ia}dx^{i}\wedge
\delta y^{a},
\end{equation*}%
which is similar to the expression in \cite{Miron-Rad}.

\item If $A_{i}=0$ (\textit{no electromagnetic potential}), then $\theta $
is the Hilbert 1-form of the space, 
\begin{equation*}
\theta =y_{i}dx^{i}
\end{equation*}%
and $\omega ,$ the fundamental 2-form of $(M,\mathcal{F}),\ $\cite{Szilasi}: 
\begin{equation*}
\omega =g_{ia}\delta y^{a}\wedge dx^{i}.
\end{equation*}
\end{enumerate}

\bigskip

\begin{definition}
We call \textit{electromagnetic tensor} in the Finslerian space, $(M,%
\mathcal{F})$, the following 2-form on $TM:$ 
\begin{equation*}
F=\omega +g_{ia}dx^{i}\wedge \delta y^{a},
\end{equation*}
\end{definition}

The above definition is equivalent to 
\begin{equation}
F=dA.  \label{F-A}
\end{equation}%
In local coordinates, we have 
\begin{equation}
F:=\dfrac{1}{2}F_{ij}dx^{i}\wedge dx^{j}+\tilde{F}_{ia}dx^{i}\wedge \delta
y^{a},  \label{def_F}
\end{equation}%
where 
\begin{equation}
F_{ij}=A_{j|i}-A_{i|j},~\ \tilde{F}_{ia}=-A_{i\cdot a},~\tilde{F}%
_{ai}=A_{i\cdot a}.  \label{F}
\end{equation}%
In relation (\ref{F}) we denoted indices corresponding to vertical geometric
objects by different letters $a,b,c...,$ in order to point out the
antisymmetry of $F.$

The above is a natural generalization of the electromagnetic tensor, for
anisotropic Finslerian spaces. The new component, $\tilde{F}_{ia},$\ will
play an important role in the equations of motion of charged particles (see
Section \ref{Lorentz_force}).

\textbf{Remark: }The electromagnetic tensor $F$ remains invariant under
transformations 
\begin{equation*}
A(x,y)~\mapsto A(x,y)+d\lambda (x),
\end{equation*}%
where $\lambda :M\rightarrow \mathbb{R}$ is a scalar function, since $%
d(A+d\lambda )=dA+d(d\lambda )=dA.$

\textbf{Particular case: }If $A=A(x)$ does not depend on the directional
variables, we get $\tilde{F}_{ia}=0$ and 
\begin{equation*}
F=\dfrac{1}{2}(A_{j|i}-A_{i|j})dx^{i}\wedge dx^{j},
\end{equation*}%
which is similar to the expression in \cite{Lagrange}, \cite{Miron-Rad}.

\section{\label{Lorentz_force}Lorentz force}

The equations of motion of a charged particle in an electromagnetic field
can be obtained from the variational procedure applied to the Lagrangian (%
\ref{L}). The corresponding Euler-Lagrange equations $\dfrac{\partial 
\mathcal{L}}{\partial x^{i}}-\dfrac{d}{dt}(\dfrac{\partial \mathcal{L}}{%
\partial y^{i}})=0$ lead to%
\begin{equation}
g_{kh}(\dfrac{dy^{h}}{dt}+2G^{i})+\dfrac{q}{c}(\dfrac{\partial A_{k}}{%
\partial x^{h}}-\dfrac{\partial A_{h}}{\partial x^{k}})y^{h}+\dfrac{q}{c}%
A_{k\cdot h}\dfrac{dy^{h}}{dt}=0,~\ \ y^{i}=\dot{x}^{i}.
\label{Lorentz-rough}
\end{equation}

Writing the second term above, in terms of covariant derivatives and taking
into account (\ref{F}) and the equality $\dfrac{\delta y^{i}}{dt}=\dfrac{%
dy^{i}}{dt}+N_{~j}^{i}y^{j}=\dfrac{dy^{i}}{dt}+2G^{i}$, we get

\begin{proposition}
\textit{(Lorentz force law): }The extremal curves $t\mapsto
(x^{i}(t)):[0,1]\rightarrow \mathbb{R}^{4}$ of the Lagrangian (\ref{L}) are
given by 
\begin{equation}
\dfrac{\delta y^{i}}{dt}=\dfrac{q}{c}F_{~h}^{i}y^{h}+\dfrac{q}{c}\tilde{F}%
_{~a}^{i}\dfrac{\delta y^{a}}{dt},  \label{Lorentz1}
\end{equation}
\end{proposition}

An elegant equivalent writing of the above can be\emph{\ }obtained in terms
of the gravito-electromagnetic\ 2-form $\omega $ in (\ref{omega}). In order
to obtain this, let us write%
\begin{equation*}
\omega =\dfrac{1}{2}\omega _{ij}dx^{i}\wedge dx^{j}+\tilde{\omega}%
_{ia}dx^{i}\wedge \delta y^{a},
\end{equation*}%
where $\omega _{ij}=\dfrac{q}{c}F_{ij},~\ \ \tilde{\omega}_{ia}=(\dfrac{q}{c}%
\tilde{F}_{ia}-g_{ia}).$ We are led to

\begin{proposition}
The equations of motion of a charged particle in electromagnetic field in
Finslerian spaces are 
\begin{equation}
\omega _{ij}(x,y)\dfrac{dx^{j}}{dt}+\tilde{\omega}_{ia}(x,y)\dfrac{\delta
y^{a}}{dt}=0,~\ y=\dot{x}.
\end{equation}
\end{proposition}

\begin{remark}
\begin{enumerate}
\item In the case of an anisotropic potential $A,$ there appears an
additional term%
\begin{equation}
\tilde{F}^{i}(x,y):=\tilde{F}_{~a}^{i}\dfrac{\delta y^{a}}{dt}=-g^{ik}(x,y)%
\tilde{F}_{ka}(x,y)\dfrac{\delta y^{a}}{dt}  \label{new_force}
\end{equation}%
in the equations of motion, in comparison to the isotropic case.

\item Both the "traditional" Lorentz force (given by $F^{i}=F_{~h}^{i}y^{h}$%
) and the correction $\tilde{F}$ are orthogonal to the velocity 4-vector $y=%
\dot{x}:$%
\begin{equation}
g_{ij}F^{i}y^{j}=0,g_{ij}\tilde{F}^{i}y^{j}=0.  \label{ortho}
\end{equation}

\item The above defined $F_{kh},~\tilde{F}_{ka},F^{\iota },\tilde{F}^{i}$
are components of \textit{distinguished tensor fields}, \cite{Lagrange}.
\end{enumerate}
\end{remark}

\bigskip

\textbf{Physical interpretation: }The usual interpretation of the extremal
curves are the equations of motion. Therefore, the expression in the right
hand side of (\ref{Lorentz1}) presents the Lorentz force in anisotropic
spaces. We see that its first term which is common with the isotropic case
is proportional to velocity, while the second term is proportional to
acceleration which brings to mind the idea of an \textquotedblright
inertial\ force\textquotedblright\ in accelerated reference frames.

\section{Homogeneous Maxwell equations}

Taking into account that $F=dA,$ we immediately get the identity $%
dF=d(dA)=0. $ In other words:

\begin{proposition}
There holds the generalized homogeneous Maxwell equation: 
\begin{equation}
dF=0,  \label{Max1}
\end{equation}
where $F$ is the electromagnetic tensor (\ref{def_F}), (\ref{F}), and $d$
denotes exterior derivative.
\end{proposition}

In local coordinates, the homogeneous Maxwell equation is read as: 
\begin{eqnarray*}
&&F_{ij|k}+F_{ki|j}+F_{jk|i}=-\underset{(i,j,k)}{\sum }R_{~jk}^{b}\tilde{F}%
_{ib}; \\
&&\tilde{F}_{aj|k}+\tilde{F}_{ka|j}+F_{jk\cdot a}=0 \\
&&\tilde{F}_{ka\cdot b}+\tilde{F}_{bk\cdot a}=0.
\end{eqnarray*}

The first set in the above is the analogue (in adapted coordinates) of the
regular homogeneous (sourceless) Maxwell equations.

There also hold the relations%
\begin{equation}
\tilde{F}_{ia}y^{i}=0,~\tilde{F}_{ia}y^{a}=0,  \label{rel_F}
\end{equation}%
entailed by the 1-homogeneity of $L_{1}$ and the fact that $A_{i}=\dfrac{%
\partial L_{1}}{\partial y^{i}}$ are its $y$-partial derivatives.

\bigskip

Conversely, on a topologically "nice enough" domain, we have

\begin{theorem}
\label{th1}If on a contractible subset of $T\mathbb{R}^{4}$ we define the
electromagnetic tensor as a 2-form%
\begin{equation*}
F:=\dfrac{1}{2}F_{ij}dx^{i}\wedge dx^{j}+\tilde{F}_{ia}dx^{i}\wedge \delta
y^{a},
\end{equation*}%
on the respective subset, satisfying%
\begin{equation*}
dF=0;
\end{equation*}%
then there exists a \textit{horizontal} 1-form%
\begin{equation*}
A=A_{i}(x,y)dx^{i}
\end{equation*}%
such that $F=dA.$ Moreover, if $\tilde{F}_{ia}y^{i}=0$ and $\tilde{F}%
_{ia}y^{a}=0,$ then $A_{i}=\dfrac{\partial L_{1}}{\partial y^{i}}$ for some
1-homogeneous in $y$ scalar function $L_{1}(x,y).$
\end{theorem}

\textbf{Proof: }By Poincar\'{e} lemma, we deduce that there exists a 1-form%
\begin{equation*}
\bar{A}=\phi _{i}(x,y)dx^{i}+\psi _{a}(x,y)\delta y^{a}
\end{equation*}%
such that $F=d\bar{A}.$ By computing $d\bar{A}$ and equating its components
with those of $F,$ we get%
\begin{equation*}
F_{ij}=\phi _{j|i}-\phi _{i|j}-R_{~ij}^{a}\psi _{a};~\ \ \tilde{F}%
_{ia}=-\phi _{i\cdot a}-\psi _{a\cdot i},~~\ 0=\psi _{a\cdot b}-\psi
_{b\cdot a}.
\end{equation*}

From the last relation, we get that there exists a scalar function $\psi
=\psi (x,y)$ such that $\psi _{a}=\psi _{\cdot a},$ $a=\overline{1,4}.$
Then, by direct computation, it follows that, if we build the following 
\textit{horizontal} 1-form:%
\begin{equation*}
A:=A_{i}dx^{i},~\ \ \ A_{i}:=\phi _{i}+\delta _{i}\psi ,
\end{equation*}%
then our 2-form $F$ is none but its exterior differential: $F=dA.$

Further, from $\tilde{F}_{ia}y^{i}=0,$ we get $A_{i\cdot a}y^{i}=0,$ which
is, $(A_{i}y^{i})_{\cdot a}=A_{\cdot a}.$ By setting $L_{1}=A_{i}y^{i}$ and
re-denoting indices, we have $A_{i}=\dfrac{\partial L_{1}}{\partial y^{i}}.$
1-homogeneity of $L_{1}$ now follows from $\tilde{F}_{ia}y^{a}=0,$ q.e.d.

\section{Currents in Finslerian spaces}

In the classical Riemannian case, \textit{the inhomogeneous Maxwell equation}
$d(\ast F)=4\pi \ast J$ can be obtained by means of the variational
principle applied to $\int (\alpha F_{ij}F^{ij}-\beta J^{k}A_{k})\sqrt{-g}%
d\Omega ,$ \cite{RBS}, where $J$ denotes the 4-vector of a current, $\alpha $
and $\beta $ are constants, $g=\det (g_{ij})$ and $d\Omega =dx^{1}\wedge
dx^{2}\wedge dx^{3}\wedge dx^{4},$ and the integral is taken on some domain
in $M=\mathbb{R}^{4}.$

In our case, the quantities $F_{ij},$ $F^{ij},$ $A_{k}$ depend on $y,$ hence
the integrand is actually defined on some domain in $TM.$ It is natural to
look for a generalization of the above Lagrangian to $TM.$ Also, it is
reasonable to think of the current as of a vector field on $TM:$%
\begin{equation}
\mathcal{J}=J^{i}(x,y)\delta _{i}+\tilde{J}^{a}(x,y)\dot{\partial}_{a}.
\label{def_J}
\end{equation}%
The meaning of the quantities $\tilde{J}^{a}$\ will reveal itself later.

\bigskip

Let us consider $A_{i}=A_{i}(x,y)$ and the following integral of action on
some domain in $TM:$

\begin{equation}
I=\int (\alpha (F_{ij}F^{ij}+\tilde{F}_{ia}\tilde{F}^{ia}+\tilde{F}_{ai}%
\tilde{F}^{ai}-\beta J^{k}A_{k})\sqrt{G}d\Omega ,  \label{I}
\end{equation}%
where $d\Omega =~dx^{1}\wedge ...dx^{4}\wedge \delta y^{1}\wedge ..\delta
y^{4},$ $G=\det (G_{\alpha \beta }),$ and $G_{\alpha \beta }$ denotes the
Sasaki lift of $g$.

In order to make physical sense for the above, we need to adjust measurement
units so as to have $[F_{ij}]=[\tilde{F}_{ia}].$ Hence, let 
\begin{equation*}
u^{a}=\dfrac{1}{H}y^{a},
\end{equation*}%
be the fibre coordinates on $TM,$ where, $H$ is a constant (ex: $[H]=\dfrac{1%
}{\sec }$) meant to have the same measurement units for $x^{i}$ and $u^{a}:$ 
$[x^{i}]=[u^{a}],$ (consequently, also $[F_{ij}(x,u)]=[\tilde{F}_{ia}(x,u)]$%
). Also, let $\dot{\partial}_{a}=\dfrac{\partial }{\partial u^{a}}.$ The
integral (\ref{I}) only involves the horizontal part $h\mathcal{J=}%
J^{i}(x,u)\delta _{i}$ of the extended current $\mathcal{J},$ let us denote
it by 
\begin{equation*}
J=J^{i}(x,u)\delta _{i}.
\end{equation*}

The integral of action (\ref{I}) is $I=\int (\alpha (F_{ij}F^{ij}+2\tilde{F}%
_{ia}\tilde{F}^{ia})-\beta J^{k}A_{k})\sqrt{G}d\Omega .$ By varying the
potentials $A_{k},$ we get 
\begin{equation*}
\mathbf{\delta }_{A}I=\int \{4\alpha (\delta A_{i;j}F^{ji}-\delta A_{i\cdot
a}\tilde{F}^{ia})-\beta J^{k}\mathbf{\delta }A_{k})\}\sqrt{G}d\Omega ,
\end{equation*}%
which is 
\begin{equation*}
\begin{array}{c}
\mathbf{\delta }_{A}I=\int 4\alpha \{(\mathbf{\delta }A_{i}F^{ji}\sqrt{G}%
)_{;j}-(\mathbf{\delta }A_{i}\tilde{F}^{ia}\sqrt{G})_{\cdot a}-(F^{ji}\sqrt{G%
})_{;j}\mathbf{\delta }A_{i}+(\tilde{F}^{ia}\sqrt{G})_{\cdot a}\mathbf{%
\delta }A_{i}\} \\ 
-\beta J^{i}\mathbf{\delta }A_{i}\sqrt{G}d\Omega .%
\end{array}%
\end{equation*}

We have\ $(\delta A_{i}F^{ji}\sqrt{G})_{;j}=div(\delta A_{i}F^{ji}\sqrt{G}%
)+\delta A_{i}F^{ji}\sqrt{G}N_{j\cdot a}^{a},~\ (\delta A_{i}\tilde{F}^{ia}%
\sqrt{G})_{\cdot a}=div(\delta A_{i}\tilde{F}^{ia}\sqrt{G}).$ The
divergences can be transformed into integrals on the boundary if the domain
of integration; by considering variations $\mathbf{\delta }A_{i}$ which
vanish on this boundary, it remains 
\begin{equation*}
\delta _{A}I=\int \left\{ 4\alpha \left( (F^{ij}\sqrt{G})_{;j}-F^{ij}\sqrt{G}%
N_{j\cdot a}^{a}+(\tilde{F}^{ia}\sqrt{G})_{\cdot a}\right) -\beta J^{i}\sqrt{%
G}\right\} \mathbf{\delta }A_{i}d\Omega =0.
\end{equation*}

\begin{proposition}
There holds the generalized inhomogeneous Maxwell equation: 
\begin{equation}
\dfrac{1}{\sqrt{G}}\{(F^{ij}\sqrt{G})_{;j}-F^{ij}N_{j\cdot a}^{a}\sqrt{G}\}+%
\dfrac{1}{\sqrt{G}}(\tilde{F}^{ia}\sqrt{G})_{\cdot a}=\dfrac{\beta }{4\alpha 
}J^{i},  \label{Max_currents}
\end{equation}
\end{proposition}

\textbf{Particular case:} If the space is pseudo-Riemannian, then $%
A_{i}=A_{i}(x)$ and, \cite{Shen}, \cite{Lagrange}, $N_{~j}^{a}=\gamma
_{~jk}^{a}(x)y^{k}$ (where $\gamma _{~jk}^{a}$ denote the Christoffel
symbols of the metric $g$), hence $N_{~j\cdot a}^{a}=\gamma _{~ja}^{a}.$ We
get%
\begin{eqnarray*}
&&\dfrac{1}{\sqrt{G}}\{(F^{ij}\sqrt{G})_{;j}-F^{ij}N_{j\cdot a}^{a}\sqrt{G}%
\}=\dfrac{1}{\sqrt{G}}\{(F^{ij}\sqrt{G})_{,~j}-(N_{~j}^{a}F^{ij}\sqrt{G}%
)_{\cdot a}\}= \\
&&\dfrac{1}{\sqrt{G}}\{F_{~~,~j}^{ij}\sqrt{G}+F^{ij}(\sqrt{G})_{,~j}-\gamma
_{~ja}^{a}F^{ij}\sqrt{G}\}= \\
&&~=F_{~~,~j}^{ij}+2F^{ij}\gamma _{~ja}^{a}-F^{ij}\gamma _{~ja}^{a}=\nabla
_{j}F^{ij}.
\end{eqnarray*}%
(where we have taken into account that $G=g^{2}$ and $(\sqrt{-g}%
)_{,i}=\gamma _{~ia}^{a}$). That is, if the space is isotropic, equations (%
\ref{Max_currents}) are just the usual ones:%
\begin{equation*}
\nabla _{j}F^{ij}=\dfrac{\beta }{4\alpha }J^{i}.
\end{equation*}

\textbf{Conclusion: }In comparison to the case of isotropic spaces, there
appears a new term in the expression of the current, namely, 
\begin{equation}
\zeta ^{i}=\dfrac{1}{\sqrt{G}}(\tilde{F}^{ia}\sqrt{G})_{\cdot a}.
\label{new-current}
\end{equation}

This means that in an anisotropic space the measured fields would correspond
to an effective current consisting of two terms: one is the current provided
by the experimental environment, the other is the current corresponding to
the anisotropy of space. The presence of the current $\zeta ^{i}$ in
experimental situations could be noticed if $\left\vert (\tilde{F}^{ia}\sqrt{%
G})_{\cdot a}\right\vert \approx \left\vert (F^{ij}\sqrt{G})_{;j}\right\vert 
$. Particularly, if the space is isotropic, then $A_{i}=A_{i}(x),$ and $%
\zeta ^{k}=0.$

\bigskip

Relation (\ref{Max_currents}) above does not involve the vertical components 
$\tilde{J}^{a}$ of the current. Hence, for the moment we have no reason to
suppose they are nonzero. Still, a formal approach using exterior derivative
would emphasize them, and they appear as necessary as "compensating"
quantities in order to obtain the continuity equation.

\bigskip

If we formally generalize inhomogeneous Maxwell equation as 
\begin{equation}
d(\ast F)=\dfrac{\beta }{4\alpha }\ast \mathcal{J},  \label{Max2}
\end{equation}%
where $\ast $ denotes the Hodge star operator on the manifold $TM$, then we
obtain by direct computation

\begin{eqnarray*}
\dfrac{1}{\sqrt{G}}\{(F^{ij}\sqrt{G})_{;j}+(\tilde{F}^{ia}\sqrt{G})_{\cdot
a}\}-F^{ij}N_{j\cdot a}^{a} &=&\dfrac{\beta }{4\alpha }J^{i} \\
\dfrac{1}{\sqrt{G}}(\tilde{F}^{ai}\sqrt{G})_{;i} &=&\dfrac{\beta }{4\alpha }%
\tilde{J}^{a},
\end{eqnarray*}%
where $\mathcal{J}=J^{i}\delta _{i}+\tilde{J}^{a}\dot{\partial}_{a}$ is as
in (\ref{def_J}).

The first set of equations is nothing but (\ref{Max_currents}) obtained by
means of variational methods, while the second one is new. We notice the
appearance of the vertical components $\tilde{J}^{a}=\dfrac{4\alpha }{\beta }%
\dfrac{1}{\sqrt{G}}(\tilde{F}^{ai}\sqrt{G})_{;i}.$

With the above expression of $\mathcal{J},$ there holds the generalized
continuity equation: $d(\ast \mathcal{J})=\dfrac{4\alpha }{\beta }d(d(\ast
F))=0,$ which is,%
\begin{equation*}
div\mathcal{J}=0.
\end{equation*}

We notice that the divergence $div(J^{i}\delta _{i})$\ of the horizontal
current is not equal to zero. In order to have charge conservation $div%
\mathcal{J}=0$, the new quantity (formally introduced) $v\mathcal{J}=\tilde{J%
}^{a}\dot{\partial}_{a}$\ is needed.

\bigskip

\textbf{Comparison to existent results:}

A previous approach for the equations of electromagnetism in anisotropic
spaces was made by R. Miron and collaborators, \cite{Lagrange}, \cite%
{Miron-Rad}, where the definition of the electromagnetic tensor is made by
means of deflection tensors of metrical linear connections. There, it is
proposed an \textit{internal electromagnetic tensor }(with \textit{h-h} and 
\textit{v-v} components), of a Lagrange space $(M,g),$%
\begin{equation}
F=F_{ij}dx^{j}\wedge dx^{i}+f_{ab}\delta y^{a}\wedge \delta y^{b},
\label{inner_F}
\end{equation}%
where $F_{ij}=\dfrac{1}{2}(y_{j|i}-y_{i|j}),$ $f_{ab}=\dfrac{1}{2}%
(y_{a}|_{b}-y_{b}|_{a})$ are defined by means of covariant derivatives
attached to a certain (metrical) linear connection\textit{\ }$D\Gamma (N)$.

In the respective works, only position dependent potentials $A(x)$ are
considered, leading to $F=F_{ij}dx^{j}\wedge dx^{i}$ (and $f_{ab}=0$).

Here, we propose an alternative definition of the electromagnetic tensor (%
\ref{def_F}),(\ref{F}) (with horizontal $hh$- and mixed\ $hv$- components
instead of $hh$- and $vv$- ones as in (\ref{inner_F})), based on the idea
that in anisotropic spaces, the electromagnetic potential is generally
direction dependent, which corresponds to the physically testable situation.
Maxwell equations are obtained here as solutions of a variational problem
and in terms of exterior derivatives, and they differ from the ones obtained
for (\ref{inner_F}). Moreover, the new components $\tilde{F}_{ia}$ of our
electromagnetic tensor have precise physical meanings, since they are
tightly related to Lorentz force. Also, newly appearing currents have a
precise role in making continuity equation fulfilled.

An analogue of (\ref{inner_F}) is obtained if we consider the Lorentz
nonlinear connection, \cite{Ingarden}, \cite{Lagrange}, of coefficients $%
\bar{N}_{~j}^{i}=N_{~j}^{i}-\dfrac{q}{c}F_{~j}^{i}$ and the linear
connection given by $\bar{D}\Gamma (N)=(\bar{L}_{~jk}^{i},\bar{C}%
_{~jk}^{i}=-g^{il}A_{l\cdot jk}),$ where $\bar{L}_{~jk}^{i}=\dfrac{1}{2}%
g^{ih}(g_{hj;k}+g_{hk;j}-g_{jk;h}).$ Then, $F$ can be described by 
\begin{equation*}
F_{ij}=\dfrac{1}{2}(y_{j||i}-y_{i||j}),~\ \tilde{F}_{ia}=g_{ia}-y_{i}||_{a}
\end{equation*}%
(where $_{||i},~||_{a}$ denote the associated covariant derivations).

\textbf{Conclusions:}

In the present paper we show that anisotropic electromagnetic potentials
lead to additional terms in the equations of motion of charged particles.
Starting from these, we build a generalization of the electromagnetic
tensor, which leads to extra homogeneous Maxwell equations and additional
terms in the expression of currents.

\textbf{Acknowledgment: }The work was supported by the grant No. 4 /
03.06.2009, between the Romanian Academy and Politehnica University of
Bucharest and the RFBR grant No. 07-01-91681-RA\_a.

\end{document}